\documentclass[a4paper]{article}

\usepackage{INTERSPEECH2021}
\usepackage{amsmath, amssymb}
\usepackage{bbold}
\usepackage{mathtools}
\title{Speech Representation Learning Combining Conformer CPC with \\Deep Cluster for the ZeroSpeech Challenge 2021}
\name{Takashi Maekaku$^1$, Xuankai Chang$^2$, Yuya Fujita$^1$, Li-Wei Chen$^2$, \\Shinji Watanabe$^2$, Alexander Rudnicky$^2$}
\address{
  $^1$Yahoo Japan Corporation, Tokyo, JAPAN\\
  $^2$Carnegie Mellon University, PA, USA}
\email{\{tmaekaku,yuyfujit\}@yahoo-corp.jp, \{xuankaic,liweiche,alex.rudnicky,swatanab\}@andrew.cmu.edu}

\begin{document}

\setlength{\abovedisplayskip}{2pt}
\setlength{\belowdisplayskip}{2pt}
\setlength\textfloatsep{2pt}
\setlength\floatsep{2pt}

\maketitle
\begin{abstract}
We present a system for the Zero Resource Speech Challenge 2021, which combines a Contrastive Predictive Coding (CPC) with deep cluster. 
In deep cluster, we first prepare pseudo-labels obtained by clustering the outputs of a CPC network with k-means.
Then, we train an additional autoregressive model to classify the previously obtained pseudo-labels in a supervised manner.
Phoneme discriminative representation is achieved by executing the second-round clustering with the outputs of the final layer of the autoregressive model.
We show that replacing a Transformer layer with a Conformer layer leads to a further gain in a lexical metric.
Experimental results show that a relative improvement of 35\% in a phonetic metric, 1.5\% in the lexical metric, and 2.3\% in a syntactic metric are achieved compared to a baseline method of CPC-small which is trained on LibriSpeech 460h data.
We achieve top results in this challenge with the syntactic metric.
\end{abstract}

\vspace{-9pt}

\section{Introduction}
\renewcommand{\thefootnote}{\fnsymbol{footnote}}

\footnote[0]{We have found that Table 2 in the previous version contains a typo that the numbers in the 5th and 6th columns are swapped. This version has corrected this typo and modified the corresponding descriptions in Section 4.2.1. Note that these changes do not affect the main logic and discussions of this paper.}
Many studies have shown that textual information is essential for building speech recognition systems and language models (LM).
Recently, several important studies on representation learning \cite{baevski2019vq,baevski2020wav2vec, chung2020generative,pascual2019learning,liu2020mockingjay} and semi-supervised training \cite{xu2020iterative,park2020improved,masumura2020sequence} explored using a large amount of speech data without corresponding text annotations and demonstrated significant improvements in speech recognition performance. 
This suggests that such systems may learn to train their own LM from raw audio only.
Therefore, it is hoped that eventually spoken language modeling tasks can be done without any text annotations.

The Zero Resource Speech (ZeroSpeech) Challenge 2021 \cite{nguyen2020zero} is designed to tackle such unsupervised LM training using only raw speech data as input.
The evaluation is done using a suite of 4 black-box, zero-shot metrics, which probe for the quality of the training models at 4 linguistic levels: phonetics, lexicon, syntax and semantics.
The baseline system consists of three components: an acoustic model, a clustering module (k-means), and an LM.
The acoustic model is built upon Contrastive Predictive Coding (CPC) \cite{oord2018representation}, where the representation of the audio is learned by predicting the future frames using an autoregressive model.
After training the CPC model, the baseline system trains a k-means clustering module on the outputs of the final layer of the autoregressive model to obtain sequences of discretized audio files.
Finally, the LM is trained with the discretized units as pseudo-labels. 

In this challenge, the final goal is to solve a couple of discrimination tasks. 
However, the representation obtained by the CPC model does not have sufficient linguistically discriminative characteristics since the CPC model itself is trained for the prediction task.
To address this issue, we propose a method that combines the CPC model with a deep cluster method \cite{caron2018deep,xie2016unsupervised,guo2017deep,hsu2020hubert}.
We train an autoregressive model for phoneme classification using pseudo-labels obtained by clustering the outputs of a CPC network using k-means. 
The phoneme discriminative representation is obtained by doing a second-round clustering on the outputs of the final layer of the autoregressive model.
Note that we call it phoneme classification in the sense of classifying pseudo-labels, which are likely to capture the phonetic meaning.

Furthermore, we examine replacing the Transformer \cite{vaswani2017attention} layer of the CPC model with a Conformer \cite{gulati2020conformer} layer.
Conformer incorporates a convolutional neural network (CNN) \cite{lecun1995convolutional} inside the Transformer to handle not only global but also local contexts, and its usefulness has been recognized in speech recognition tasks \cite{gulati2020conformer, zhang2020pushing,huang2020improving,guo2020recent}.
Likewise, it is expected that more precise phonetic and lexical representation is achieved by capturing both contexts using the Conformer network.
We apply the above two methods separately and confirm that both methods outperform the baseline method using the phonetic metric.
In addition, we observe that the proposed method combining the Conformer CPC model with the deep cluster method outperforms the baseline method using  the lexical metric.
This reveals that the two methods have a complementary effect on both tasks.





\section{Challenge Overview}
\vspace{-3pt}
\label{sec:overview}
In this section, we briefly introduce the baseline system and the task of the ZeroSpeech Challenge 2021 \cite{nguyen2020zero}.

\vspace{-3pt}
\subsection{Baseline System}
\label{ssec:baseline}
\begin{figure}[tbp]
  \includegraphics[clip,width=8.2cm]{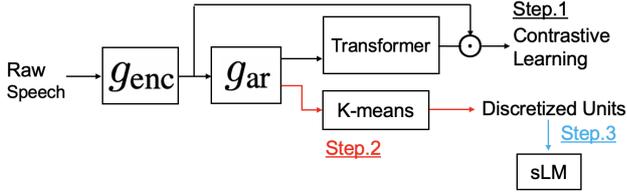}
  \caption{\textrm Illustration of the baseline system. First, we train a Contrastive Predictive Coding (CPC) model which consists of $\mathit{g}_\text{enc}$ and $\mathit{g}_\text{ar}$ optimized by Eq.~(\ref{eq:contrastive_loss}) (Step.1). Then, k-means clustering is performed to generate discretized units of audio data (Step.2). Finally, we train a spoken language model (sLM) using the discretized units as pseudo-labels (Step.3).}
  \label{fig:baseline}
\end{figure}

The baseline system consists of a speech representation learning model, a clustering model, and a language model.
Fig~\ref{fig:baseline} illustrates the architecture of the baseline system.

\subsubsection{Contrastive Predictive Coding}
\label{sssec:cpc}

The speech representation model is based on CPC, a self-supervised representation learning method proposed in \cite{oord2018representation}. Instead of using a conditional generative model to predict the future input signal, the CPC model learns the representation via maximizing the mutual information between the current context and future embeddings. The CPC model consists of two modules. First, given an input speech signal $\mathbf{x}$, a non-linear encoder $\mathit{g}_\text{enc}$ maps it to a $T$-length sequence of embeddings with a lower time resolution: $\mathbf{z} = \mathit{g}_\text{enc} (\mathbf{x}) $, where $\mathbf{z} = (z_1, \dots, z_T)$. Then, an autoregressive encoder $\mathit{g}_\text{ar}$ aggregates the information from $\mathbf{z}$, producing a context latent representation $c_t = \mathit{g}_\text{ar}(\mathbf{z}_{\leq t}), t\in\{1,\dots,T\}$. The CPC model is optimzied by minimizing the noise-contrastive estimation-based (NCE) loss \cite{gutmann2010noise}. At each time $t$, given the context representaion $c_t$ and its $K$ future embeddings $\{z_{t+k}\}_{1 \leq k \leq K}$, the loss is defined as:
\begin{align}
    \mathcal{L}_t = - \frac{1}{K} \sum_{k=1}^K \log \left[ \frac{\exp(z_{t+k}^T h_k(c_t))}{ \sum_{\tilde z \in \mathcal N_t} \exp(\tilde {z}^T h_k(c_t))} \right], \label{eq:contrastive_loss}
\end{align}
where $\mathcal{N}_t$ is a set of negative embedding samples and $h_k(\cdot)$ is a transformation for each step $k$.
In this challenge, we use two different versions of CPC model: CPC-small and CPC-big, the differences of which are elaborated in Table~\ref{tab:cpc-params}.

\vspace{-3pt}
\subsubsection{Clustering and Language Models}
\label{ssssec:lm}

To train a spoken language model (sLM) on pseudo-labels, the raw speech signal needs to be mapped to a sequence of discrete symbols. The pre-trained CPC model first generates a sequence of representations given the raw speech signal as input. Then, these representations are used to train a clustering model, which is k-means, with 50 clusters used in this work. 

After training, the clustering model is applied to the speech representation of the training data to produce class labels. The class label can be regarded as a pseudo linguistic subword unit. Using these label sequences as pseudo-text data, we can train an sLM.
In this work, we trained a BERT \cite{vaswani2017attention} language models. This model consists of multiple Transformer layers.
Note that the BERT model is only trained with the masked language model objective, following \cite{liu2019roberta}. Finally, the score of the language model on the pseudo-label sequence is regarded as a pseudo-probability (PP).

\subsection{Dataset}
\label{ssec:dataset}

The training data is comprised of the audio from the LibriSpeech 960h dataset \cite{panayotov2015librispeech} and the Libri-light dataset \cite{kahn2020libri}. The CPC-small model is trained on the 100 hours of clean audio subset (train-clean-100) from the LibriSpeech data, while the CPC-big model is trained on a 6K-hour subset of Libri-light data. 
The k-means clustering is performed on the train-clean-100h subset to obtain the centroid coordinates.
Then the k-means estimates the pseudo-label sequences on LibriSpeech 960h data, which becomes the training set for the language model.

Each of the four metrics is evaluated on its \textit{dev} and \textit{test} sets, which are specially designed for the corresponding task. Please refer to the challenge description \cite{nguyen2020zero} for more details of how the evaluation data are generated.

\vspace{-6pt}
\subsection{Evaluation Metrics}
\vspace{-5pt}
The performance of the spoken language model is evaluated using four different metrics, each corresponding to a task at a specific linguistic level: 
phonetics, lexicon, syntax and semantics.

\textbf{Phonetics.}~
The ABX metric \cite{schatz2013evaluating} discriminates the speech sound between phonetic minimal pairs (e.g. ``aba" and ``apa"). Given the speech sounds $a$, $x$ and $b$, where $a$ and $b$ are from two categories $A$ and $B$ ($A \neq B$), and $x$ belongs to category $A$ respectively, it computes the probability that the two sounds from the same category are closer than the two sounds from different categories:
\begin{align}
    \hat e (A,B) \coloneqq &\frac{1}{n_{A}(n_A -1)n_B} \sum_{\substack{a,x\in A \\ x \neq a}}\sum_{b \in B}\left[ \mathbb{1}_{d(b,x)<d(a,x)} + \right. \nonumber \\
    & \left. \frac{1}{2} \mathbb{1}_{d(b,x)=d(a,x)} \right].
    \label{eq:abx}
\end{align}
$n_{A} \text{ and } n_{B}$ represent the cardinalities of category $A$ and $B$.


\textbf{Lexicon.}~
The sWUGGY ``Spot-the-word" \cite{le2017comparing} is used to discriminate an existing word from a lexically similar non-word using the sLM (e.g. ``brick" and ``blick"). The metric measures the accuracy that the PP of the real word is higher than that of the non-word: $\mathbb{1}_{\text{PP}(\text{word}) > \text{PP}(\text{non-word})}$. 

\textbf{Syntax.}~
sBLIMP acceptability, adapted from BLIMP \cite{warstadt2020blimp}, discriminates a grammatical sentence from an ungrammatical sentence (e.g. ``dogs eat meat" and ``dogs eats meat"). The metric accepts it if the PP of a grammatical sentence is greater than the ungrammatical one: $\mathbb{1}_{\text{PP}(\text{Sentence}_\text{gram}) > \text{PP}(\text{Sentence}_\text{ungram})}$.

\textbf{Semantic.}~
sSIMI similarity measures the similarity between the representations of pairs of words and compares the results with human judgment. The metric is computed as the Spearman's rank correlation coefficient $\rho$ between the semantic similarity scores given by the model and the human scores in the dataset.

\setlength{\tabcolsep}{10pt}
\begin{table*}[th]
  \caption{Characteristics of the baseline acoustic CPC models. We took the last LSTM layer of CPC-small and the second LSTM hidden layer of CPC-big as inputs to the clustering.}
  \label{tab:cpc-params}
  \centering
  \begin{tabular}{  l c c c c }
    \toprule
    \multicolumn{1}{c}{\textbf{Model}} & \multicolumn{2}{c}{\textbf{CPC model configuration}} &  \multicolumn{1}{c}{\textbf{Training data}} &                          \multicolumn{1}{c}{\textbf{Input to k-means}} \\ \cline{2-3} 
    & autoregressive & hidden units & &  \\ 
    \toprule
    CPC-small                       & $2$-layer LSTM & $256$ & LibriSpeech clean-100h & 2nd layer of LSTM~~~             \\
    CPC-big                       & $4$-layer LSTM & $512$ & Libri-light clean-6kh & 2nd layer of LSTM~~~             \\

    \bottomrule
  \end{tabular}
  
\end{table*}


  

\vspace{-3pt}
\section{Proposed System}

The two proposed methods are described below. As each of these methods modifies a separate component in the baseline system, they can be used in combination.

\subsection{CPC with deep cluster}
\begin{figure}[ht]
  \centering{\includegraphics[clip,width=8.2cm]{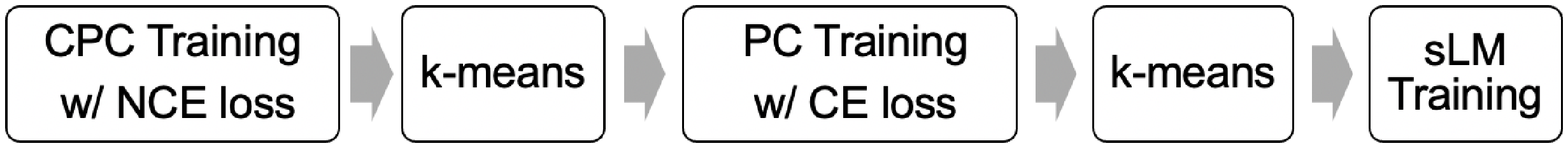}}
  \caption{\textrm Illustration of our proposed system (CPC with deep cluster). First, we train a CPC model which consists of $\mathit{g}_\text{enc}$ and $\mathit{g}_\text{ar}$ optimized by Eq.~(\ref{eq:contrastive_loss}). Then, k-means clustering is performed to generate discretized units of audio data. Next, another CPC network is trained for phoneme classification (PC) using the discretized units as pseudo-labels. After that, we obtain more linguistically discriminative representation by second-round clustering. Finally, we train a sLM based on pseudo-labels.}
  \label{fig:cpc_dc}
\end{figure}

All four evaluation metrics in this challenge are discriminative tasks. However, as we mentioned, the baseline system does not have sufficiently linguistically discriminative characteristics.
To solve this problem, our system combines the CPC model with the deep cluster method \cite{caron2018deep,xie2016unsupervised,guo2017deep}.
Deep cluster is a clustering method initially designed for image processing. It iterates between doing k-means clustering on the features produced by a neural network and updating its weights by classifying the cluster assignments of each feature.
HUBERT \cite{hsu2020hubert} is similar to our method in that it uses the deep cluster method to perform self-learning.
Fig~\ref{fig:cpc_dc} illustrates the architecture of our method. First, we follow the same procedure as the baseline system until the k-means clustering step. After that, we obtain the discretized pseudo-labels for each feature frame.
Then, we randomly initialize\footnote{In a preliminary experiment, we compared the case where the network is initialized with the first round of the CPC network weights and the case where the network is reinitialized randomly. As a result, better performance was obtained with the latter.} a new model with the same architecture as the original one.
However, this time the objective is to classify pseudo-labels of feature steps with the cross-entropy (CE) criterion, which is more straightforward than the NCE loss.

Finally, we execute the second-round k-means clustering with the outputs of the final layer of the autoregressive model. A phoneme discriminative representation is achieved by imposing a phoneme classification task with the pseudo-labels on the autoregressive model.

\vspace{-2pt}

\subsection{Conformer CPC}
\vspace{-3pt}
We propose Conformer CPC which replaces the Transformer classifier $h_k(\cdot)$ in Eq.~(\ref{eq:contrastive_loss}) with a Conformer block.
It contains two Feed Forward modules sandwiching the Multi-Headed Self-Attention \cite{vaswani2017attention} module and the Convolution module.
For input $c$ in Eq.~(\ref{eq:contrastive_loss}) to a Conformer block, the output $y$ of the block is:


\begin{align}
\label{conformer}
  & \tilde{c} = c + \frac{1}{2}\mathrm{FFN}(c),\\
  & c' = \tilde{c} + \mathrm{MHSA}(\tilde{c}),\\
  & c'' = c' + \mathrm{Conv}(c'),\\
  & y = \mathrm{Layernorm}(c''+\frac{1}{2}\mathrm{FFN}(c''))
\end{align}
where FFN refers to the Feed Forward module, MHSA refers to the Multi-Head Self-Attention module, and Conv refers to the Convolution module as described in \cite{gulati2020conformer}.
This network can capture not only long-term contexts via the self-attention block but also local contexts through a Convolution module.
Therefore, it is expected that more precise phonetic and lexical representations are achieved.



\label{ssec:dc}

\begin{table*}[th]
  \caption{Within (all stimuli $a$, $b$ and $x$ in Eq.~(\ref{eq:abx}) are uttered by the same speaker) and Across ($a$ and $b$ are from the same speaker, and $x$ from a different speaker) Speaker ABX metric (lower is better) on Libri-light dev-clean and dev-other. All embeddings are extracted from the final layer of the autoregressive network before clustering.
  ``DC" stands for the deep cluster.
  ``1st" of Training Data means a data set for contrastive learning and ``2nd" of that means a data set for phoneme classification. Each model is trained on LibriSpeech (LS) or Libri-light (LL).}
  \label{tab:abx}
  \centering
  \begin{tabular}{  l c c c c c c}
    \toprule
    \multicolumn{1}{c}{\textbf{Embedding}} & \multicolumn{2}{c}{\textbf{Training Data}} & \multicolumn{2}{c}{\textbf{within (\(\downarrow\))}} & \multicolumn{2}{c}{\textbf{across (\(\downarrow\))}} \\ \cline{2-3} \cline{4-5} \cline{6-7} 
    & 1st & 2nd & dev-clean & dev-other & dev-clean & dev-other  \\ 
    \toprule
    Baseline : CPC-small      & LS-100h    & /        & 6.24 & 8.48 & 8.17 & 13.55             \\
    Baseline : CPC-small       & LS-460h       & /      & 6.19	&  8.71 & 7.34 & 13.02             \\
    Proposed: Conformer CPC-small        & LS-100h    & /         & 5.78 &  8.23 & 7.83 & 13.59             \\
    Proposed: Conformer CPC-small       & LS-460h & / & 5.40 &  7.55 & 7.17 & 12.19             \\
    Proposed: CPC-small+DC          & LS-100h  & LS-100h & 4.78 &  7.01 & 6.78 & 12.34             \\
    Proposed: CPC-small+DC  & LS-460h & LS-460h & \textbf{3.93} 	&  \textbf{5.99} & \textbf{5.18} & \textbf{10.00}  \\
    Proposed: Conformer CPC-small+DC  & LS-460h & LS-460h & 4.05 &  6.12 & 5.38 &	10.60  \\
    \hline
    Baseline : CPC-big           & LL-6kh & /         & 3.41 &  \textbf{4.85} & 4.18 & \textbf{7.64}             \\

    Proposed: CPC-big+DC       & LL-6kh & LS-960h        & 3.28	&  4.96	& 4.14	& 8.28             \\
    Proposed: CPC-big+DC (1024units)      & LL-6kh & LS-960h & \textbf{3.11} & 	4.96 & \textbf{3.98} & 7.92             \\

    \bottomrule
  \end{tabular}
  
\end{table*}

\begin{table*}[th]
  \caption{Overall performance (higher is better) of the baseline and the proposed models on dev sets on three zero-shot metrics. For all models, the k-means clustering (k=50) was performed on LibriSpeech clean-100h, and the BERT-small models were trained on discretized units of LibriSpeech 960h.}
  \label{tab:other_metrics}
  \centering
  \begin{tabular}{  l c c c c c c}
    \toprule
    \multicolumn{1}{c}{\textbf{System}} & \multicolumn{2}{c}{\textbf{Training Data}} &  \multicolumn{1}{c}{\textbf{sWUGGY (\(\uparrow\))}} & \multicolumn{1}{c}{\textbf{sBLIMP (}\(\uparrow\))} & \multicolumn{2}{c}{\textbf{sSIMI (\(\uparrow\))}} \\
    \cline{2-3}  \cline{6-7} 
    & 1st & 2nd & & & synth. & libri.  \\ 
    \toprule
    Baseline : CPC-small     & LS-100h & /   & 65.79 & 52.88 & -0.09 & \textbf{9.23}
		    \\
    Baseline : CPC-small     & LS-460h & /    & 66.21 & 52.79 & -0.67 & 4.92
		    \\
    Proposed: Conformer CPC-small     & LS-100h & /    & 62.22 & 52.96 & \textbf{0.90} & 7.22 \\
    Proposed: Conformer CPC-small     & LS-460h & /    & 66.10 & \textbf{53.39} & -1.84 & 5.17
			 \\
    Proposed: CPC-small+DC     & LS-100h & LS-100h    & 65.42 & 52.86 & -1.10 & 8.14
		    \\
    Proposed: CPC-small+DC     & LS-460h & LS-460h    & 64.89 & 52.75 & -2.11 & 8.89   \\
    Proposed: Conformer CPC-small+DC     & LS-460h & LS-460h    & \textbf{67.21} & 53.38 &	-0.17 & 7.07
		   \\
	\hline
    Baseline : CPC-big     & LL-6kh & /          & 65.81 &	52.91 & \textbf{3.88} & \textbf{5.56}    \\

    Proposed: CPC-big+DC    & LL-6kh & LS-960h         & \textbf{66.01} & \textbf{54.15} & -0.81 & 5.45
			    \\
	Proposed: CPC-big+DC (1024units)    & LL-6kh & LS-960h         & 62.64 & 54.06	& -1.65 &4.81
			    \\

    \bottomrule
    \vspace{-30pt}
  \end{tabular}
  
\end{table*}

\section{Experiments}

\subsection{Experimental Setup}
Following the baseline system \cite{nguyen2020zero}, the encoder $\mathit{g}_\text{enc}$ consists of five 1d-convolutional layers with kernel sizes of (10, 8, 4, 4, 4) and stride sizes of (5, 4, 2, 2, 2). The downsampling factor of $\mathit{g}_\text{enc}$ is 160 and the embedding $\mathbf{z}$ has a sampling rate of 100Hz. Then, the multi-layer long short-term memory (LSTM) \cite{hochreiter1997long} network is used as an autoregressive encoder $\mathit{g}_\text{ar}$. The CPC model can be divided into two categories: CPC-small and CPC-big, the differences of which are elaborated in Table~\ref{tab:cpc-params}.

The transformation $h_k(\cdot)$ in Eq.~(\ref{eq:contrastive_loss}) is a 1-layer Transformer or Conformer network, the parameters of which are as follows:
The number of attention heads is $8$ and the hidden unit size is $512$.
The number of hidden units for the feed-forward layers is $2048$.
As for Conformer, the kernel size of the convolution module is $30$.
During the training of CPC models, we applied dropout \cite{srivastava2014dropout} with a rate of $0.1$ for the Transformer and Conformer block in the same way as existing studies \cite{vaswani2017attention,gulati2020conformer} to achieve a better generalization. We also applied dropout with a rate of $0.5$ for the outputs of the CPC prediction network before taking the product with $z$.
$K$ in (\ref{eq:contrastive_loss}) was set to $12$.

The number of iterations for k-means clustering was set to $150$.
This is the same for the first-round clustering and the second-round one. The language model was based on BERT \cite{liu2019roberta}. We reduced the number of parameters by considering the training time. The model consists of $8$ Transformer layers, each of which has $8$ attention heads with hidden dimensionality of $512$. The dimensionality of feed-forward layers is $2048$. The sLM can be trained within 60 hours on a single GPU using the pseudo-text of LibriSpeech 960h.

All models were implemented with PyTorch, including CPC\_audio\footnote{https://github.com/facebookresearch/CPC\_audio} and fairseq\footnote{https://github.com/pytorch/fairseq}.
The former was a modified version of the CPC that stabilizes the CPC training by replacing batch normalization \cite{ioffe2015batch} with a channel-wise normalization.
The latter was only used for the sLM training.

We identified three baseline systems: CPC-small trained on LibriSpeech 100h and 460h, respectively, and CPC-big trained on Libri-light 6kh.
Seven proposed systems that combine different methods and training data sizes were included.
The proposed methods do not necessarily require the same configuration for the initial autoregressive network and the network for phoneme classification.
For this reason, we also compared a system in which the size of the hidden units in the network for phoneme classification was increased from $512$ to $1024$.

\vspace{-3pt}
\subsection{Results and Discussion}
\vspace{-3pt}

\subsubsection{ABX metric}
In Table~\ref{tab:abx}, we present the results of the ABX metric for the baseline system and our two proposed systems before clustering.
It is clear that almost all proposed systems of CPC-small outperform the original CPC-small baseline.
The combination of CPC with deep cluster and Conformer CPC improves the performance up to 35\% relative to the baseline, although not as much as the performance of CPC with deep cluster alone.
This shows that the two proposed systems yield linguistically discriminative characteristics for the CPC network.
Comparing the CPC-big models, we see that our systems outperform the baseline system only under the condition of ``dev-clean".
One possible reason for this is that the training data for the phoneme classification task in the 2nd stage was LibriSpeech 960h and was not sufficient compared with the baseline CPC-big training with Libri-light 6kh.

\vspace{-6pt}
\subsubsection{sWUGGY metric}
\vspace{-6pt}
Table~\ref{tab:other_metrics} compares sWUGGY, sBLIMP, and sSIMI metrics with
the baseline, and the proposed methods.
The two proposed systems, when applied independently, failed to outperform the baseline results compared to the CPC-small baseline systems.
Therefore, better performance in the ABX metric does not necessarily guarantee better performance in the sWUGGY metric.
However, the best performance is achieved when the two proposed systems are applied simultaneously, i.e., Conformer CPC-small+DC.
Compared to the CPC-small trained on LibriSpeech 460h data, Conformer CPC-small+DC achieves a relative improvement of 1.5\%.
This result suggests that the two methods have a complementary effect on the lexical metric.
\vspace{-3pt}
\subsubsection{sBLIMP metric}
\vspace{-3pt}
The proposed system of CPC-big with deep cluster achieves the highest score among all methods in Table~\ref{tab:other_metrics}.
This is also the top result in this challenge\footnote{The leader-board can be viewed at https:\slash\slash zerospeech.com\slash 2021\slash results.html.}.
Besides, we can see that all Conformer CPC systems outperform all baseline systems regardless of the amount of training data.
It indicates that the Conformer block works to help learn higher-level linguistic features.
\vspace{-6pt}
\subsubsection{sSIMI metric}
\vspace{-3pt}
For all methods including the proposed systems and the baseline systems, there are no systems that are significantly better for both synthetic (synth.) and LibriSpeech (libri.) sets.
We can see that the amount of training data does not directly contribute to the performance improvement even if comparing within baseline methods.
The proposed systems generally achieve a performance that is almost competitive with the baseline systems.
\vspace{-8pt}

\section{Conclusions}
\vspace{-3pt}
In this paper, we have proposed a system which combines CPC with deep cluster.
In deep cluster, we first prepare pseudo-labels obtained by clustering the outputs of a CPC network with k-means. Then, we train an additional autoregressive classifier to predict the previously obtained pseudo labels in a supervised manner.
Phoneme discriminative representation is achieved by executing the second-round clustering with the outputs of the final layer of the autoregressive model.
In addition, we show that replacing the Transformer layer with a Conformer layer leads to a further gain in a lexical metric.
Experimental results show that a relative improvement of 35\% in a phonetic metric, 1.5\% in the lexical metric, and 2.3\% in a syntactic metric are achieved compared to a baseline method of CPC-small which is trained on LibriSpeech 460h data.
This result suggests that both methods have a complementary effect on the lexical metric.

\newpage
\bibliographystyle{IEEEtran}

\bibliography{mybib}

\end{document}